\documentclass[prb,twocolumn,showpacs,amsmath,amssymb]{revtex4}

\usepackage{graphicx}
\usepackage{amssymb,amsmath}
\usepackage{dcolumn}
\usepackage{bm}
\usepackage{color}

\begin{document}

\title{Reply to "Comment on 'Topological stability of the half-vortices in spinor exciton-polariton condensates'"}

\author{H. Flayac}
\affiliation{Clermont Universit\'{e}, Universit\'{e} Blaise Pascal, LASMEA, BP10448, 63000 Clermont-Ferrand, France}
\affiliation{CNRS, UMR6602, LASMEA, 63177 Aubi\`{e}re, France}

\author{I.A. Shelykh}
\affiliation{Science Institute, University of Iceland, Dunhagi-3,IS-107, Reykjavik, Iceland}
\affiliation{International Institute for Physics, UFRN - Universidade Federal do Rio	Grande do Norte, Campus Universitario Lagoa Nova, CEP: 59078-970, Natal- RN, Brazil}

\author{D.D. Solnyshkov}
\affiliation{Clermont Universit\'{e}, Universit\'{e} Blaise Pascal, LASMEA, BP10448, 63000 Clermont-Ferrand, France}
\affiliation{CNRS, UMR6602, LASMEA, 63177 Aubi\`{e}re, France}

\author{G. Malpuech}
\affiliation{Clermont Universit\'{e}, Universit\'{e} Blaise Pascal, LASMEA, BP10448, 63000 Clermont-Ferrand, France}
\affiliation{CNRS, UMR6602, LASMEA, 63177 Aubi\`{e}re, France}

\begin{abstract}
In a recent work [H. Flayac, I.A. Shelykh, D.D. Solnyshkov and G. Malpuech, \emph{Phys. Rev. B} \textbf{81}, 045318 (2010)], we have analyzed the effect of the TE-TM splitting on the stability of the exciton-polariton vortex states. We considered classical vortex solutions having cylindrical symmetry and we found that the so-called half vortex states[Yu. G. Rubo, \emph{Phys. Rev. Lett.} \textbf{99}, 106401 (2007)] are not solutions of the stationary Gross-Pitaevskii equation. In their Comment [M. Toledo Solano, Yu.G. Rubo, \emph{Phys. Rev. B} \textbf{82}, 127301 (2010)], M. Toledo Solano and Yuri G. Rubo claim that this conclusion is misleading and pretend to demonstrate the existence of static half-vortices in an exciton-polariton condensate in the presence of TE-TM splitting. In this reply we explain why this assertion is not demonstrated satisfactorily.
\end{abstract}


\maketitle

First of all, we now agree with the authors of the Comment\cite{Comment} that our claim of the non-stability of half-vortices\cite{Rubo} in the presence of TE-TM splitting is not supported enough. Indeed, in our paper\cite{Hugo} we considered only solutions of the type ${\psi _ \pm }\left( {r,\phi } \right) = {f_ \pm }\left( r \right)\exp \left( {i{\theta _ \pm }\left( \phi  \right)} \right)$, where the amplitudes $f_\pm$ of the wave function depend only on $r$ and phases $\theta_\pm$ are linear with respect to $\phi$. From this point of view the publication of a comment seems reasonable.

On the other hand  Solano and Rubo considered only
asymptotic equations. They found nontrivial half-vortex asymptotic solutions and we agree with their result. In the other limit $\rightarrow0$, the kinetic energy dominates all other terms, and a cylindrically symmetric half-vortex solution can be found. However, in the intermediate region (close to the vortex core), all three contributions of the Hamiltonian (kinetic energy, interactions energy, and TE-TM splitting) are of the same order.
The existence of the asymptotic solutions close to infinity and close to the ground state does not guarantee the existence of a half-vortex solution in the whole space. As we show below, the amplitude of these asymptotic solutions depends on both $r$, and $\phi$. The general solution is therefore of the type ${\psi_ \pm }\left( {r,\phi } \right) = {f_ \pm }\left( {r,\phi } \right)\exp \left( {i{\theta _ \pm }\left( {r,\phi } \right)} \right)$. The analysis of the existence, or non-existence in the whole space of such type of half vortex is a very demanding task, which has not been performed neither by the authors of the Comment, nor by the authors of the reply. Qualitatively, we hardly believe that such vortices, with particles moving on non-circular trajectories could exist as their symmetry differs from that of the Hamiltonian and therefore some preferential directions appear. The authors of the Comment believe the opposite. At this stage, there is still no definitive answer given to the question of the stability of half vortices in the presence of TE-TM splitting.

The asymptotic equation was obtained by the authors of the Comment by variation of the elastic energy of the vortex neglecting completely the radial dependence. We believe that this way important terms could be lost. The starting point should not be the elastic energy (which is not the total energy), but the Gross-Pitaevskii equations, which in the dimensionless form read (see Eqs. 11, 17 of Ref.\onlinecite{Hugo}):

\begin{widetext}
\begin{eqnarray}
\underbrace {-\frac{1}{{2{r^2}}}\frac{{{\partial ^2}{\psi _ \pm }}}{{\partial {\phi ^2}}} + \chi {e^{ \mp 2i\phi }}\left[ { \mp \frac{{2i}}{{{r^2}}}\frac{\partial }{{\partial \phi }} + \frac{1}{{{r^2}}}\frac{{{\partial ^2}}}{{\partial {\phi ^2}}}} \right]{\psi _ \mp }}_1 + \underbrace {\left( {|{\psi _ \pm }{|^2} - 1} \right){\psi _ \pm }}_2\\
\nonumber
-\underbrace {\frac{1}{2}\left[ {\frac{{{\partial ^2}}}{{\partial {r^2}}} + \frac{1}{r}\frac{\partial }{{\partial r}}} \right]{\psi _ \pm } + \chi {e^{ \mp 2i\phi }}\left[ { - \frac{{{\partial ^2}}}{{\partial {r^2}}} \pm \frac{{2i}}{r}\frac{{{\partial ^2}}}{{\partial r\partial \phi }} + \frac{1}{r}\frac{\partial }{{\partial r}}} \right]{\psi _ \mp }}_3 = 0\end{eqnarray}
\end{widetext}
where according to the notations of the Comment we used $\phi$ for the polar angle and for simplicity neglected the interactions of polaritons with opposite circular polarizations, $\alpha_2=0$. We divided all the terms into three groups: those containing derivatives by $\phi$ only (group 1), those without derivatives (group 2) and those containing derivatives by $r$ only (group 3). Considering the asymptotic case, one sees that the terms of the group 1 are of the order of magnitude of $r^{-2}$, while the terms of the group 3 are of the order $o(r^{-2})$ and thus can be neglected, as it was done by the authors of the Comment. As for the terms of the group 2, it can be easily shown that they are of the same order of magnitude as terms from the group 1.  Indeed, using the approximative solution for the radial function of a vortex with a winding number 1, Eq. 23 of Ref.\onlinecite{Hugo} one easily sees

\begin{equation}
|\psi^2|-1\approx\frac{r}{\sqrt{1+r^2}}-1\sim\frac{1}{r^2}, r\rightarrow+\infty
\end{equation}

Thus, the asymptotic equation reads:

\begin{eqnarray}
\left[-\frac{1}{2}\frac{d^2}{d\phi^2}+\lambda_\pm\right]\psi_\pm
+\chi e^{\mp2i\phi}\left[\mp2i\frac{d}{d\phi}
+\frac{d^2}{d\phi^2}\right]\psi_\mp=0
\label{AsymptoticEq}
\end{eqnarray}
where:
\begin{equation}
{\lambda _ \pm } = \mathop {{\rm{lim}}}\limits_{r \to  + \infty } \left[ {{r^2}\left( {|{\psi _ \pm }{|^2} - 1} \right)} \right]
\end{equation}

If the solution is cylindrically symmetric, $\lambda_\pm$ should be independent on $\phi$.

Now, using the ansatz of the authors of the Comment
\begin{equation}
\psi_\pm=e^{i(\theta(\phi)\mp\eta(\phi))}
\label{Ansatz}
\end{equation}
one has 4 equations for the unknown functions $\theta,\eta,\lambda_\pm$ (since both real and imaginary parts of the equation \ref{AsymptoticEq} should be equal to zero):
\begin{widetext}
\begin{eqnarray}
-\frac{1}{2}(\theta''-\eta'')+\chi \textrm{cos}[2(\eta-\phi)]\cdot (\theta''+\eta'')+\chi \textrm{sin}[2(\eta-\phi)]\cdot(\theta'+\eta')(2-\theta'-\eta')=0\label{Eq1}\\
-\frac{1}{2}(\theta''+\eta'')+\chi \textrm{cos}[2(\eta-\phi)]\cdot (\theta''-\eta'')-\chi \textrm{sin}[2(\eta-\phi)]\cdot(\theta'-\eta')(-2-\theta'+\eta')=0\label{Eq2}\\
\frac{1}{2}(\theta'-\eta')^2+\lambda_++\chi \textrm{cos}[2(\eta-\phi)]\cdot (\theta'+\eta')(2-\theta'-\eta')-\chi \textrm{sin}[2(\eta-\phi)]\cdot(\theta''+\eta'')=0\label{Eq3}\\
\frac{1}{2}(\theta'+\eta')^2+\lambda_-+\chi \textrm{cos}[2(\eta-\phi)]\cdot (\theta'-\eta')(-2-\theta'+\eta')+\chi \textrm{sin}[2(\eta-\phi)]\cdot(\theta''-\eta'')=0\label{Eq4}
\end{eqnarray}
\end{widetext}

The main equations of Toledo Solano and Rubo (Eqs. 6a and 6b of the Comment) can be obtained taking the sum and the difference of Eqs. \ref{Eq1} and \ref{Eq2} above, which ensures that the imaginary part of Eq.\ref{AsymptoticEq} is zero. However, to make the real part being zero as well, one needs to satisfy additionally Eqs. \ref{Eq3},\ref{Eq4} as well. These equations, not written in the Comment, allow to deduce the phi dependence of $\lambda_\pm$.

For solutions of the class found in Ref.\onlinecite{Hugo}, $\lambda_\pm$ are evidently constant and given by $\eta=\phi$, $\theta=p\phi$, $\theta''=\eta''=0$, $\eta'=1,\theta'=p$ with $p$ being an integer number and

\begin{equation}
\lambda_\pm=-\frac{(p\mp1)^2}{2}+\chi(p^2-1)
\end{equation}

On the other hand, if $\theta$ and $\eta$ are complicated non-linear functions of $\phi$, in order to satisfy Eqs.\ref{Eq3},\ref{Eq4} $\lambda_\pm$ is $\phi$ dependent, which means that the amplitudes of the vortex solutions at large but finite distance from the core region are non-cylindrical, $\phi$ dependent:

\begin{equation}
\psi_\pm\approx\left[1+\frac{\lambda_\pm(\phi)}{2r^2}\right]e^{i(\theta(\phi)\mp\eta(\phi))}
\end{equation}
where the dependence of $\lambda_\pm$ on $\phi$ is deduced from Eqs.\ref{Eq3},\ref{Eq4}. As one clearly sees, in these solutions the rotational symmetry present in the initial Hamiltonian is broken and preferential directions appear. The peculiar choice of the author of the Comment is that the polariton pseudo-spin is aligned with the TE-TM effective field at $\phi=0$. But any other direction could be chosen equivalently. The existence of the solutions of this type in the whole space is not \emph{a priori} excluded, but is neither guaranteed.

Regarding the qualitative remark of Toledo Solano and Rubo that singularities in Gross-Pitaevskii equation can not appear or disappear we would like to mention that first of all, for a multi-component (e.g. spinor) condensate, in contrast with a scalar one, the conservation of current circulation and, consequently, vortex stability, is not in general guaranteed by topological considerations. In such a system a vortex can be removed by a continuous transformation \cite{Leggett}, which has been demonstrated experimentally\cite{Matthews}. Also it was already shown that vortices and vortex-antivortex pairs can appear in Gross-Pitaevskii equation from regular initial conditions, e.g. in a process of scattering of a BEC on localized defect \cite{Frisch} or in the rotating condensates \cite{Kasamatsu}. It was also shown that TE-TM splitting can generate vortices of a winding number 2 from regular initial distributions of the polariton condensate\cite{Tim}.

To conclude, we believe that the claimed stability of half-vortices under the effect of TE-TM splitting was not fully demonstrated. We disagree with the remark of the authors of the Comment that our article contains mathematical error as we found the exact vortex solutions possessing the symmetry of the Hamiltonian. The existence of half-vortices in the presence of TE-TM splitting is not proven by the authors of the Comments, and this statement should not be claimed by them. On the other hand, some of the conclusions of Ref. \onlinecite{Hugo} are indeed not fully supported.  Further works will be needed in order to give a definitive answer to this problem. Finally, one of the main conclusion of our work was that under the influence of the TE-TM field specific pairs of half-vortices become bounded to form the $(-1,+1)$ integer vortex solution (this was also remarked by the authors of the Comment). As a half-vortex will never live alone inside the microcavity one should reasonably expect that such a bounding phenomenon will occur and should leave the BKT transition temperature linked with integer vortices.


\begin{thebibliography}{99}

\bibitem{Comment} M. Toledo Solano, Yu.G. Rubo, \emph{Phys. Rev. B} \textbf{82}, 127301 (2010).

\bibitem{Rubo} Yu. G. Rubo, \emph{Phys. Rev. Lett.} \textbf{99}, 106401 (2007).

\bibitem{Hugo} H. Flayac, I.A. Shelykh, D.D. Solnyshkov and G. Malpuech, \emph{Phys. Rev. B} \textbf{81}, 045318 (2010).

\bibitem{Frisch} Frisch T, Y. Pomeau, and S. Rica, Phys. Rev. Lett. 69, 1644 (1992)

\bibitem{Leggett} A.J. Leggett, Quantum Liquids, p.154 (Oxford University Press, 2008).

\bibitem{Matthews} M.R. Matthews et al, Phys. Rev. Lett \textbf{83}, 3358 (1999).

\bibitem{Kasamatsu} K. Kasamatsu et al, \emph{Phys. Rev A} 71, 063616 (2005)

\bibitem{Tim} T. Liew, A.V. Kavokin, I.A. Shelykh, \emph{Phys. Rev. B} \textbf{75}, 241301 (2007).

\end{thebibliography}
\end{document}